\documentclass[superscriptaddress,aps,prd,twocolumn,showpacs,floatfix,nofootinbib,longbibliography, superscriptaddress,showkeys]{revtex4-1}
\usepackage{multirow}
\usepackage{url}
\usepackage{hyperref}
\usepackage{color}
\usepackage{cancel}
\usepackage{soul}
\usepackage[normalem]{ulem}

\usepackage[dvipsnames]{xcolor}
\usepackage{amstext,amssymb}
\usepackage{amsmath}
\usepackage{graphicx}
\usepackage{diagbox}
\usepackage{xspace}
\usepackage{color}
\usepackage{units}
\usepackage[T1]{fontenc}
\usepackage{amsmath,bm}
\usepackage{nicefrac}
\usepackage{slashed} 
\usepackage{multirow}
\usepackage{float}
\usepackage{subcaption}

\newcommand{\be}{\begin{equation}}
\newcommand{\ee}{\end{equation}}
\newcommand{\bea}{\begin{eqnarray}}
\newcommand{\eea}{\end{eqnarray}}

\usepackage{slashed}

\newcommand{\nua}[1]{\ensuremath{\rlap{\kern-2.5pt\ensuremath{\overset{\scriptscriptstyle(-)}{\phantom{\nu}}}}{\ensuremath{{\nu}_{#1}}}}\xspace}

\definecolor{brickred}{rgb}{0.8, 0.25, 0.33}
\definecolor{brightcerulean}{rgb}{0.11, 0.67, 0.84}
\definecolor{brown(traditional)}{rgb}{0.59, 0.29, 0.0}

\begin{document}
\title{Phenomenology of different cross-section models at DUNE}

\author{Subrajit Puhan}
\email{subrajit.puhan@niser.ac.in}
\affiliation{National Institute of Science Education and Research Bhubaneswar, Khurda, Odisha 752050, India, }
\affiliation{Homi Bhabha National Institute, Training School Complex, Anushakti Nagar, Mumbai 400094, India}

\author{Monojit Ghosh}
\email{mghosh@irb.hr}
\affiliation{Center of Excellence for Advanced Materials and Sensing Devices, Ruđer Bošković Institute, 10000 Zagreb, Croatia}

\author{Rukmani Mohanta}
\email{rmsp@uohyd.ac.in}
\affiliation{School of Physics, University of Hyderabad, Hyderabad - 500046, India}

\author{Amit Pal}
\email{amit.pal@niser.ac.in}
\affiliation{National Institute of Science Education and Research Bhubaneswar, Khurda, Odisha 752050, India, }
\affiliation{Homi Bhabha National Institute, Training School Complex, Anushakti Nagar, Mumbai 400094, India}

\author{S.K.~Swain}
\email{sanjay@niser.ac.in}
\affiliation{National Institute of Science Education and Research Bhubaneswar, Khurda, Odisha 752050, India, }
\affiliation{Homi Bhabha National Institute, Training School Complex, Anushakti Nagar, Mumbai 400094, India}

\begin{abstract}
In this paper, we study the impact of different cross-section models in the measurement of the neutrino oscillation parameters in DUNE. In particular, for the quasi-elastic (QEL) region, we considered the Llewellyn-Smith formalism (LS)
and the Hartree–Fock Continuum Random Phase Approximation (HF-CRPA) and for the resonance (RES) region, we consider the Rein–Sehgal (RS)  and the Berger–Sehgal (BS) models, and compare our results with the DUNE cross-section tune (Valencia model for QEL and RS model for RES), which was considered in their technical design report. Our results show that while the DUNE tune is best for QEL, the best model for RES is BS. As the DUNE energy region is mainly dominated by RES, for the total cross-section, the HF-CRPA+BS model provides the best strength in the cross-section, whereas the DUNE tune is the weakest among all the configurations considered in our work. Regarding the neutrino mass ordering, CP violation and octant sensitivity, the HF-CRPA+BS model provides ($25- 30$)\% improvement, and regarding the precision of the $\theta_{23}$ and $\Delta m^2_{32}$, the same model provides $(15 - 20)$\% improvement as compared to the DUNE tune. 

\end{abstract}

\maketitle
\flushbottom
\clearpage

\section{INTRODUCTION}
The precise determination of the neutrino oscillation parameters has been an active area of research in particle physics over the past two decades. Future long baseline experiments, such as Hyper-Kamiokande~\cite{Abe:2018uyc} and DUNE (formerly known as LBNE)~\cite{Abi:2020loh}, are expected to reduce the uncertainties associated with these parameters, especially the CP violating phase $\delta_{\rm CP}$, the atmospheric mixing angle $\theta_{23}$, and the sign of the atmospheric mass square difference $\Delta m^2_{32}$ to answer the fundamental questions of the neutrino sector: what is the ordering of neutrino masses, does   CP violation occur in the lepton sector, what is the octant of $\theta_{23}$, can oscillation studies provide hints of new physics, etc. Since the precise measurements of these parameters depend upon the neutrino energy, we need to reconstruct the energy on an event-by-event basis, which requires proper modeling for neutrino interactions. There are several neutrino event generators available for this purpose, such as GENIE~\cite{Andreopoulos:2009rq}, NuWro~\cite{Juszczak:2005zs}, GiBUU~\cite{Buss:2011mx}, NEUT~\cite{Hayato:2021qbb}, etc. However, due to the complexity of the nuclear medium, effects such as Fermi motion, nuclear binding energy, Pauli blocking~\cite{Bodek:2021pbg}, multinucleon correlations (2p2h) and final state interactions (FSI)~\cite{Dytman:2009fs}, theoretical modeling of neutrino-nucleus scattering, and hence the reconstruction of neutrino energy from the final state particles in detectors have become a challenge~\cite{AlvarezRuso:2017oui}.

\begin{figure*}[t]
\centering
\begin{minipage}[b]{0.45\textwidth}
\centering
\includegraphics[width=0.85\textwidth]{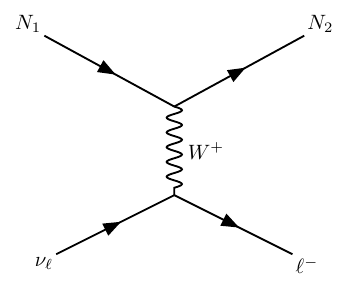}
\end{minipage}
\hfill
\begin{minipage}[b]{0.50\textwidth}
\centering
\includegraphics[width=0.85\textwidth]{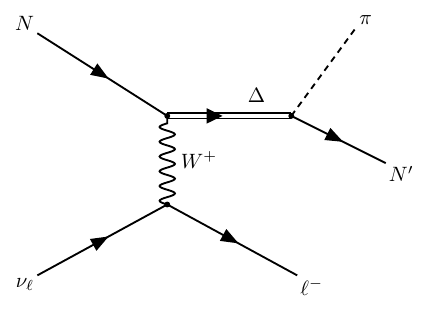}
\end{minipage}
\caption{Charged-current neutrino interactions: quasi-elastic scattering (left) and resonance production via $\Delta(1232)$ (right).}
\label{fig:cc_processes}
\end{figure*}

The neutrino energy in the range of $0.5 -5$ GeV is dominated by the quasi-elastic (QEL) and resonance (RES) interactions, where nucleons participate in single-nucleon knock-out processes and baryon excitations that decay into pions, respectively. The corresponding Feynman diagrams for both processes are shown in Fig.~\ref{fig:cc_processes}. Recent measurements also show that the contribution of 2p2h interaction in this energy region is significant and must be taken into account~\cite{PhysRevD.81.092005, PhysRevC.80.065501}. These processes are less understood because they involve complex nuclear effects, including many-body dynamics and final-state interactions, making them difficult to disentangle experimentally from one another and therefore leading to larger modeling uncertainties. These regions are especially relevant for future experiments, such as DUNE, as its beam spectrum covers a large part of the above-mentioned energy range and peaks where QEL and RES processes are dominant~\cite{Abi:2020evt}. So the accurate modeling of neutrino interaction in these regions is quite often recognized as the crucial part of the analysis. There are many theoretical models available to explain these nuclear processes, some of which are currently in use with various neutrino experiments. For a review on the role of neutrino interaction uncertainties in current and future measurements of neutrino oscillation, we refer to Ref.~\cite{Dolan:2026nlr}. In this study, without going into the details of the accuracy of the models, we investigate which cross-section model configurations lead to improved precision in the extraction of oscillation parameters in the context of the upcoming DUNE experiment. For this purpose, we use different cross-section models from GENIE. For the quasi-elastic (QEL) region, we consider several commonly used nuclear models, including Llewellyn-Smith formalism~\cite{LlewellynSmith:1972uhs} and the Hartree–Fock Continuum Random Phase Approximation (HF-CRPA)~\cite{PhysRevC.65.025501, PhysRevC.92.024606}. For the resonance (RES) region, we compare the widely used Rein–Sehgal (RS)~\cite{Rein:1980wg} and Berger–Sehgal (BS) models~\cite{PhysRevD.76.113004}. Our analysis explores the impact of these models on standard oscillation physics and compares our results with the sensitivities as given in the DUNE technical design report (TDR) \cite{DUNE:2020ypp}. We identify which combinations of QEL and RES models provide the most precise measurement of oscillation parameters, demonstrating the critical role of interaction modeling in future long-baseline neutrino experiments. Note that in the current analysis DUNE uses a different cross-section tune than the one considered in DUNE TDR \cite{DUNEModel2025}. However, in our present study, we focus on comparing our sensitivities with respect to the DUNE TDR. A comparative analysis using the current DUNE tune will be performed in a separate study.

The outline of the paper goes as follows. In the next section, we will give a detailed description of the QEL and RES theories that we consider in our work. In section~\ref{exp}, we will provide the experimental details of DUNE and the details about our numerical calculation. Section~\ref{res} contains our results, and we summarize in section~\ref{conclusion}.

\section{Cross-section models}

The cross-section models that we consider in our work are listed in table~\ref {tab:example_5x6}. In the following subsections, we discuss them in detail. 

\begin{table}[htbp]
    \centering
    \caption{Cross-section model configurations. See text for details. 
    }
    \label{tab:example_5x6}
    \begin{tabular}{lcccc}
        \hline\hline
        Configuration & Ground State & QEL & RES & DIS \\
        \hline
        DUNE Tune & BR & Valencia & RS & BY \\
        LS + RS       & BR & LS        & RS & BY \\
        LS + BS       & BR & LS        & BS & BY \\
        CRPA + RS     & BR & HF-CRPA  & RS & BY \\
        CRPA + BS     & BR & HF-CRPA  & BS & BY \\
        \hline\hline
    \end{tabular}
\end{table}

\subsection{Quasi-Elastic scattering} 

In QEL scattering, the incoming neutrinos interact with the nucleons with an exchange of $W^{\pm}$($Z$) boson via charged current (neutral current). At low momentum transfer (low $q^2$), the incoming neutrino interacts with the nucleon as a single particle. The final state involves the corresponding charged lepton and one nucleon (cf. left panel of Fig.~\ref{fig:cc_processes}). Several formalisms and approaches from the nuclear physics sector are used to explain this type of interaction. Below, we discuss the models that we will consider in this work.

\subsubsection{Valencia Model}

Although the simplest nuclear model to describe the ground state of the nucleons is the Relativistic Fermi Gas (RFG), which assumes the nucleons to be freely moving (Fermi motion) inside the nucleus, correlated by Pauli's principle, the Local Fermi Gas developed by the Valencia group~\cite{Nieves:2011pp} provides a more realistic description by incorporating the spatial dependence of the nuclear density. The local Fermi momentum is expressed as
\begin{equation}
k_F(r)=\left(\frac{3\pi^2\rho(r)}{2}\right)^{1/3},
\end{equation}
where $\rho(r)$ represents the nuclear density as a function of position inside the nucleus. The effects of Pauli blocking are incorporated through the Lindhard function, which describes the particle-hole response of the nuclear medium. The central feature of this model is the implementation of Random Phase Approximation (RPA) for the long-range correlations. These correlations are described through an effective particle-hole interaction of Landau-Migdal type, 
\begin{equation}
V_{ph}=c_0\left(f_0+f_0'\,\boldsymbol{\tau}_1\cdot\boldsymbol{\tau}_2
+g_0\,\boldsymbol{\sigma}_1\cdot\boldsymbol{\sigma}_2
+g_0'\,\boldsymbol{\sigma}_1\cdot\boldsymbol{\sigma}_2\,
\boldsymbol{\tau}_1\cdot\boldsymbol{\tau}_2\right),
\end{equation}
where $c_0$ is a normalization constant, $f_0$ and $f_0'$ represent the spin-independent isoscalar and isovector interactions, respectively, while $g_0$ and $g_0'$ correspond to the spin-dependent isoscalar and isovector interactions. Here, $\boldsymbol{\sigma}$ and $\boldsymbol{\tau}$ denote the spin and isospin Pauli matrices, respectively. The RPA series includes both particle-hole (ph) and $\Delta$-hole ($\Delta h$) excitations, providing a more complete description of the collective nuclear response. Final-state interactions (FSI) are incorporated through the nucleon self-energy, giving rise to spectral function effects that substantially broaden the quasielastic (QEL) peak. The model developed by the Valencia group was used for describing the quasi-elastic interactions in DUNE's technical design report~\cite{Abi:2020evt}. During the implementation, T2K’s 2017/8 parameterization of the Valencia RPA eﬀect was used.

\subsubsection{Llewellyn-Smith Formalism (LS)}
 
 Llewellyn-Smith Formalism (LS) describes charged-current quasi-elastic (CCQE) neutrino-nucleon scattering, mediated by a $W^\pm$ boson. The hadronic weak current is formulated using Lorentz invariant form factors reflecting the internal structure of nucleons, and the differential cross section is given as
\begingroup\small
\begin{equation*}
\frac{d\sigma}{dQ^2} = \frac{G_F^2 \ M^2\cos^2\theta_C}{8\pi E_\nu^2}
\left[ A(Q^2) \pm B(Q^2)\frac{s-u}{M^2} + C(Q^2)\frac{(s-u)^2}{M^4} \right]
\end{equation*}
\endgroup
where \[
s - u = 4 M E_\nu + Q^2 - m^2\;.
\]
$A(Q^2)$, $B(Q^2)$, and $C(Q^2)$ are the structure functions expressed as combinations of the nucleon form factors. Here, $G_F$ is the Fermi coupling constant, $\theta_C$ is the Cabibbo angle, $E_\nu$ represents the incident neutrino energy, \(M\) (\(m\)) denotes the nucleon (charged lepton) mass,  and \(Q^2\) is the squared four-momentum transfer (negative in scattering conventions). The vector and pseudoscalar form factors are determined through either experimental measurements or constrained by theoretical principles such as the Conserved Vector Current (CVC) hypothesis and the Partial Conservation of the Axial Current (PCAC). The only unknown remaining is the axial form factor $F_A(Q^2)$, which  is often modeled using a dipole parameterization of the form 
\[
F_A(q^2) = F_A(0)\left(1 + \frac{Q^2}{M_A^2} \right)^{-2},
\]
where \( M_A \) is the axial mass parameter. In GENIE, its default value is set to 0.99~GeV/\(c^2\)~\cite{Andreopoulos:2015wxa}.  The value of \( F_A(0) \) is found to be $-1.23 \pm$ 0.01 from the beta decay experiments. 

\subsubsection{Hartree--Fock continuum random phase approximation (HF-CRPA)}

The Continuum Random Phase Approximation (CRPA) provides a microscopic description of nuclear excitations and plays an important role in modeling lepton--nucleus scattering, especially in the low and intermediate energy regimes where collective nuclear effects cannot be ignored. Its development can be understood as a natural extension of the Independent Particle Model (IPM), the Hartree--Fock (HF) mean-field picture, and the standard RPA. A major advantage of CRPA is that it provides a realistic microscopic description of the nuclear response at energies where the ejection of nucleons is important. It naturally incorporates long-range correlations and collective effects while correctly treating the continuum of scattering states. This makes CRPA particularly valuable for modeling electron and neutrino scattering at low momentum transfers, where the cross section is strongly influenced by nuclear dynamics beyond the impulse approximation. 

Another approach to explain QEL scattering is by super-scaling~\cite{Donnelly:1999sw, Donnelly:1999cp}. Based on inclusive $(e,e')$ scattering data, various types of scaling behaviour have been observed~\cite{West:1975ui, Caballero:2010us, Antonov:2011xm}. The general idea is to define a dimensionless scaling function by dividing the measured cross section by an appropriate single-nucleon cross section and plotting it as a function of a suitably defined scaling variable. If the resulting function exhibits no dependence on the momentum transfer $q$, it is said to satisfy \emph{scaling of the first kind}. If it also shows no dependence on the nuclear target (i.e., the Fermi momentum $k_F$), it is referred to as \emph{scaling of the second kind}. When both conditions are met, the phenomenon is called \emph{super-scaling}. Later, this model was extended to 2p2h region (SuSAv2)~\cite{PhysRevC.90.035501}. The implementation of HF-CRPA in GENIE interpolates between HF-CRPA and SuSAv2 responses at large momentum transfer using a special version of the SuSA tensors valid up to 10 GeV. This is because its validity may decrease at high momentum transfer due to the lack of full relativistic dynamics~\cite{PhysRevD.106.073001}. That is why, when the HF–CRPA framework is employed for the QEL channel, combined with SuSA at higher energies through an interpolation, the 2p2h contribution is instead modeled using SuSAv2, as it better matches the hybrid character of this approach. Overall, CRPA represents a significant refinement over simpler nuclear models. By combining the self-consistent mean field of HF with the dynamical correlations of RPA and a proper treatment of continuum states.

\subsection{Resonance production}
\label{Sec: Form}
The energy range of 1 to 5~GeV, commonly referred to as the \textit{resonance region}, plays a crucial role in accelerator-based long-baseline neutrino experiments. In this range, the neutrino cross section is dominated by the production of baryon resonances, particularly the \( \Delta(1232) \) resonance, which subsequently decays into a nucleon and a pion (cf. right panel of Fig.~\ref{fig:cc_processes}). This process bridges the gap between quasi-elastic scattering at lower energies and deep inelastic scattering at higher energies. For the RES region, there are two widely used models, namely the Rein-Sehgal and Berger-Sehgal models.
\noindent

\subsubsection{Rein Sehgal Model (RS)}

The Rein–Sehgal (RS) model offers a unified treatment of 
neutrino-induced baryon resonance production in the few-GeV energy range. Within this approach, an incoming neutrino excites a bound nucleon to a baryon resonance $R$ of 
invariant mass $W$, which then decays (most commonly into $\pi N$ final states). The model is built upon the Feynman–Kislinger–Ravndal (FKR) framework~\cite{PhysRevD.3.2706} for baryon structure, supplemented by a non-resonant isospin-1/2 background. In the GENIE implementation, 16 of the 18 possible resonances are included, specifically those identified as unambiguous in the most recent Particle Data Group (PDG) listings. Possible interference between adjacent resonances is not taken into account. The axial transition form factor is assumed to have a dipole form,
\begin{equation}
    F_A(Q^2) = \frac{F_A(0)}{\left(1 + Q^2/M_A^2\right)^2},
\end{equation}
with $M_A =1.121~\mathrm{GeV}/c^2$, and incorporates the pion–pole term required by PCAC. The RS model is formulated in terms of relativistic harmonic-oscillator quark wave functions, treats the outgoing charged lepton as massless in its original version, and does not incorporate explicit corrections for nuclear-medium effects. Nevertheless, the influence of the charged-lepton mass on the kinematic phase-space limits is taken into account. Despite these simplifying assumptions, the model yields a computationally efficient and phenomenologically accurate description of single-pion production processes and has been widely implemented in neutrino-event generators such as GENIE, NEUT, and NuWro.

\subsubsection{Berger Sehgal Model (BS)}

The Berger–Sehgal (BS) model is an improved version of the traditional RS framework for neutrino-induced single-pion production. It was introduced to address several well-known deficiencies of the original RS model. While the BS model preserves the core theoretical structure of the RS approach, it adds three key refinements:  
(i) updated resonance parameters,  
(ii) inclusion of non-resonant background terms, and  
(iii) explicit lepton-mass corrections, which are particularly important for the $\nu_\mu$ channel.
The central improvement in the Berger–Sehgal approach is the adoption of modern electromagnetic 
helicity amplitudes and experimentally determined transition strengths from the PDG, instead of the 
older quark-model inputs used in the RS formulation. 
This leads to a substantially better description of the resonant contributions, most notably the $\Delta(1232)$.
A crucial theoretical advance is the explicit treatment of the outgoing lepton mass $m_\ell$. The RS model assumes the outgoing charged lepton to be massless, which is not a good approximation for $\nu_\mu$ interactions at energies below about 2 GeV. The Berger–Sehgal model improves this by including the lepton mass effects directly in the cross-section calculation. Because of this, the BS model keeps the simplicity of the Rein–Sehgal approach while giving more reliable predictions at lower energies.


\section{Experimental and simulation details}
\label{exp}

To simulate the DUNE experiment, we use the official GLoBES \cite{Huber:2004ka,Huber:2007ji} files corresponding to the DUNE TDR \cite{Abi:2103.04797}. DUNE will utilize a high-intensity neutrino beam generated by a 1.2 MW proton beam produced at Fermilab in Batavia, Illinois, to be detected at the far detector, to be constructed 1,300 kilometers away at the Sanford Underground Research Facility in Lead, South Dakota. The far detector will consist of four 17 kton Liquid Argon Time Projection Chamber (LArTPC) detector modules~\cite{Abi:2020evt}. In our analysis, we consider a run-time of 6.5 years in both neutrino and antineutrino modes. 

In this study, the total cross section for neutrino–nucleus interactions is computed using GENIE, incorporating various models that we discussed in the previous section to describe the interaction dynamics over a broad energy spectrum. The initial nuclear state is modeled with the Bodek–Ritchie Fermi gas (BR) approach~\cite{PhysRevD.24.1400}, which accounts for the high-momentum tails generated by short-range nucleon–nucleon correlations. The deep inelastic scattering (DIS) component is described using the Bodek–Yang (BY) formalism~\cite{Bodek:2021pbg}, while the two-particle–two-hole (2p2h) contribution is taken from the Valencia model~\cite{Gran:2013kda}. 

For the estimation of oscillation sensitivities, we follow the standard Poisson log-likelihood approach, where the test statistic is assumed to be $\chi^{2}$-distributed. The statistical contribution is defined as
\begin{equation}
\chi^{2}_{\text{stat}} = 2 \sum_{i=1}^{n}
\left[
N^{\text{test}}_{i} - N^{\text{true}}_{i}
+ N^{\text{true}}_{i} \ln
\left(
\frac{N^{\text{true}}_{i}}{N^{\text{test}}_{i}}
\right)
\right],
\label{eq:poisson_chi2}
\end{equation}
where $N^{\text{true}}_{i}$ and $N^{\text{test}}_{i}$ denote the predicted event
rates in the $i$-th energy bin for the true and test hypotheses, respectively,
and $n$ is the total number of reconstructed energy bins. Systematic uncertainties are incorporated using the pull method~\cite{PhysRevD.66.053010, Huber:2002mx}, where each uncertain quantity is associated with a nuisance
parameter constrained by a corresponding penalty term. The oscillation parameters used in the analysis correspond to the best-fit values reported by NuFIT~v6.0\footnote{NuFIT 6.0 (2024), \url{http://www.nu-fit.org}} for the normal mass ordering, as summarized in table~\ref{tab:osc-params}. The parameters $\sin^2\theta_{23}$, $\delta_{\mathrm{CP}}$ and $\Delta m^2_{32}$ are minimized in their $3 \sigma$ allowed values. 

\begin{table}[htbp]
    \centering
    \caption{Oscillation parameters used in the sensitivity analysis.}
    \label{tab:osc-params}
    \begin{tabular}{lc}
        \hline\hline
        Parameter & Value \\
        \hline
        $\sin^2\theta_{12}$ & 0.308 \\
        $\sin^2\theta_{13}$ & 0.02215 \\
        $\sin^2\theta_{23}$ & 0.470 \\
        $\Delta m^2_{21}$   & $7.49 \times 10^{-5}\,\text{eV}^2$ \\
        $\Delta m^2_{32}$   & $2.513 \times 10^{-3}\,\text{eV}^2$ \\
        $\delta_{\mathrm{CP}}$ & $212^\circ$ \\
        \hline\hline
    \end{tabular}
\end{table}

\begin{figure*}
    \centering
    \begin{minipage}[t]{0.48\textwidth}
        \centering
        \includegraphics[width=\textwidth]{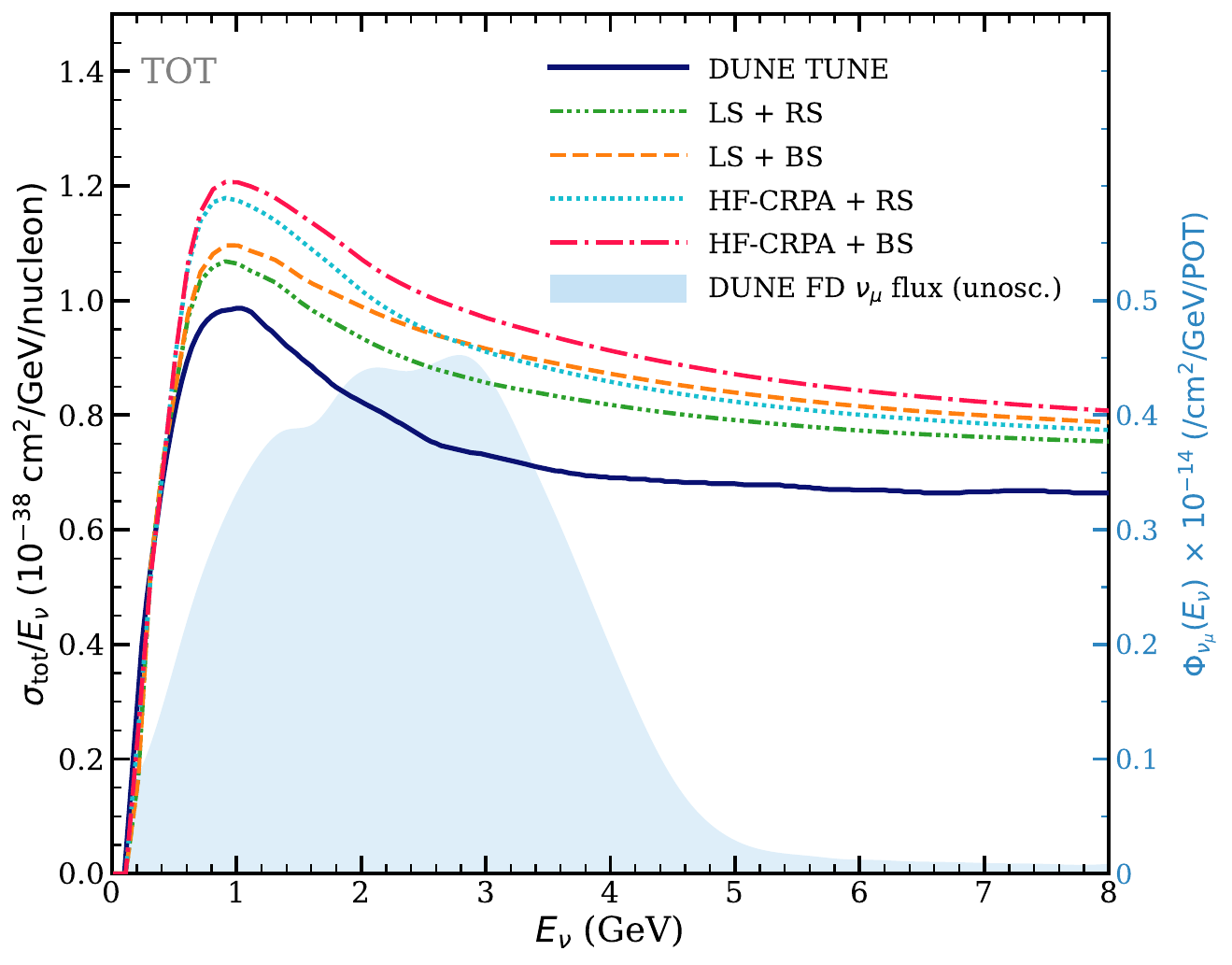}
    \end{minipage}
    \hfill
    \begin{minipage}[t]{0.48\textwidth}
        \centering
        \includegraphics[width=\textwidth]{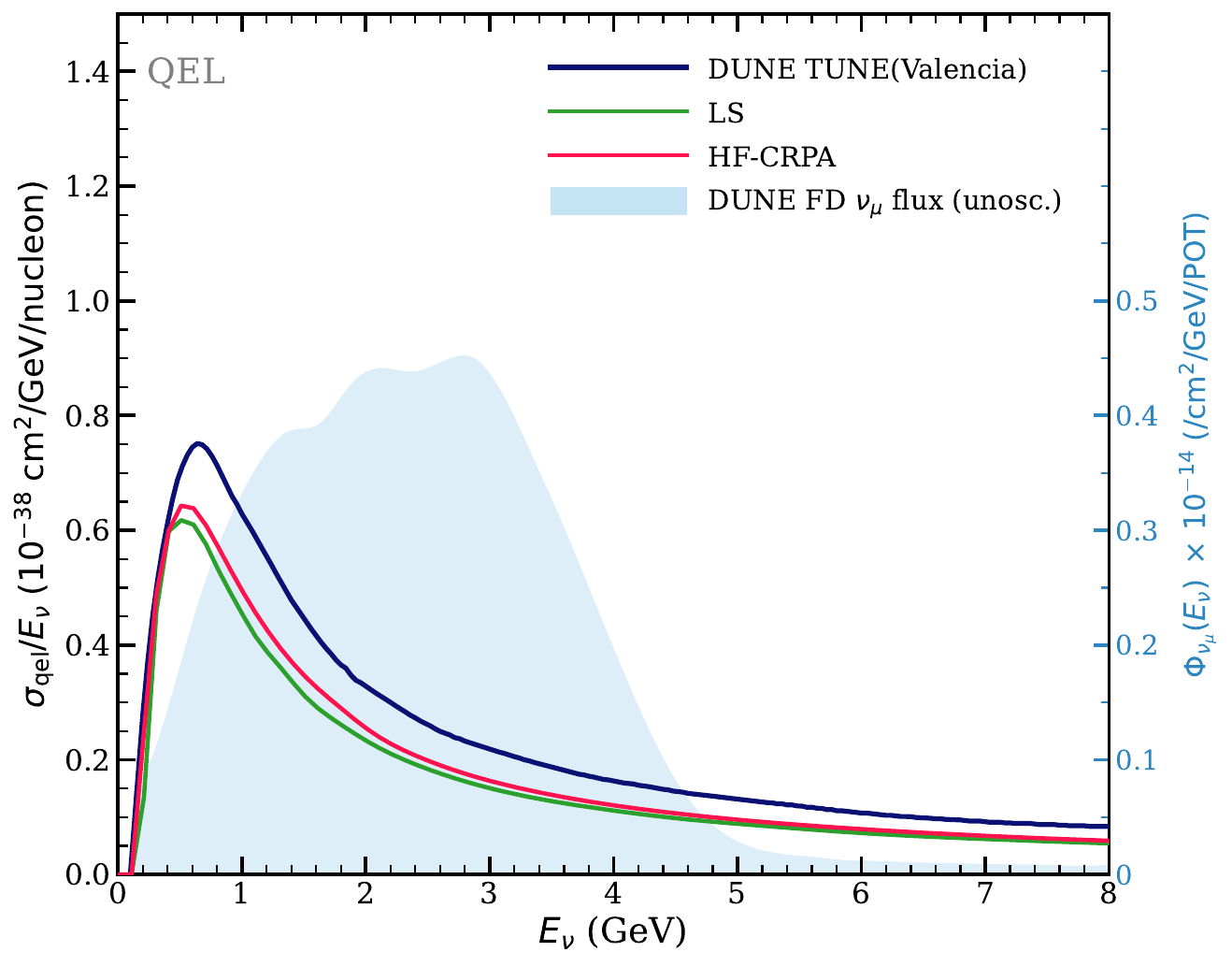}
    \end{minipage}
    \\[0.5em]
    \begin{minipage}[t]{0.48\textwidth}
        \centering
        \includegraphics[width=\textwidth]{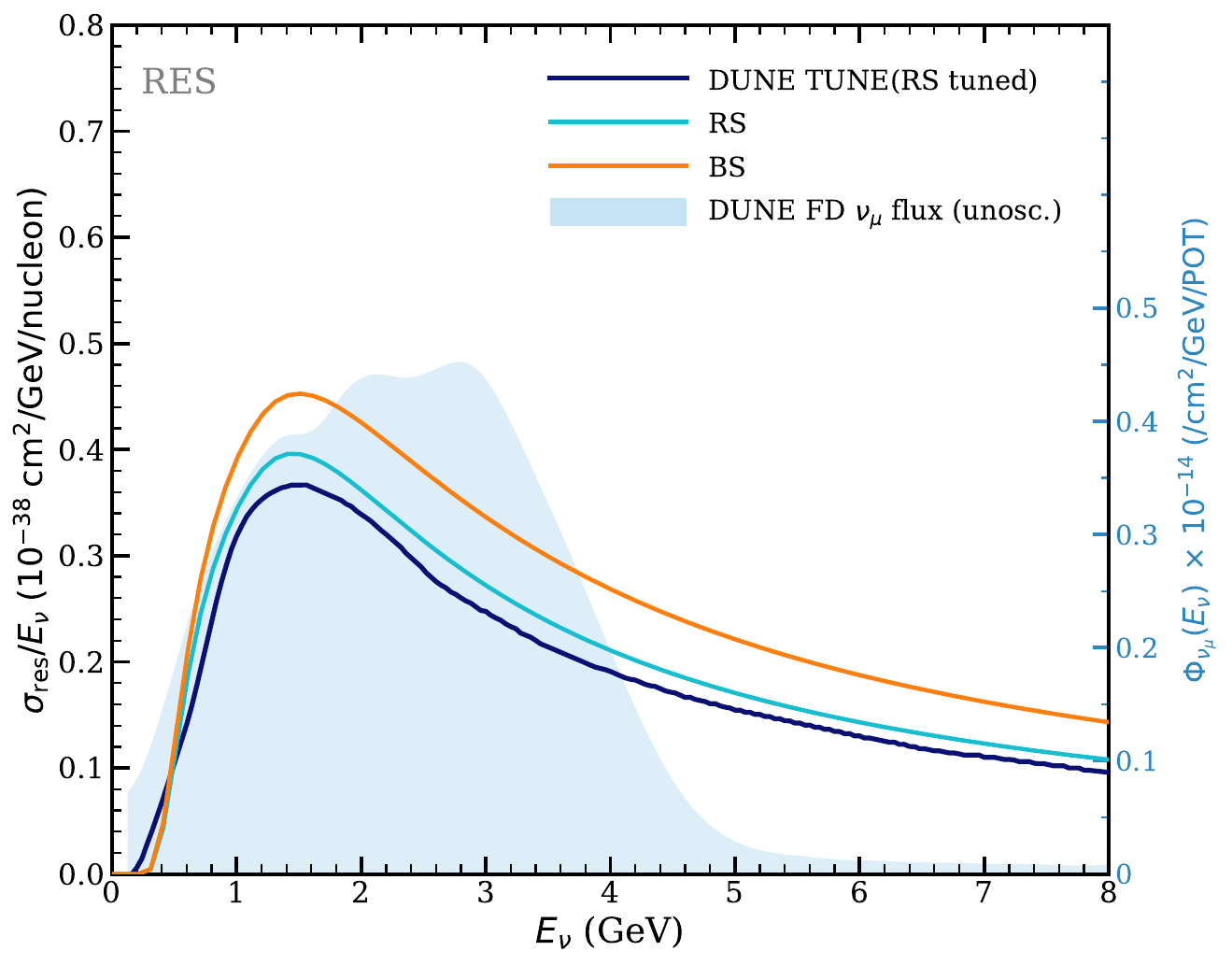}
    \end{minipage}
    \hfill
    \begin{minipage}[t]{0.48\textwidth}
        \centering
        \includegraphics[width=\textwidth]{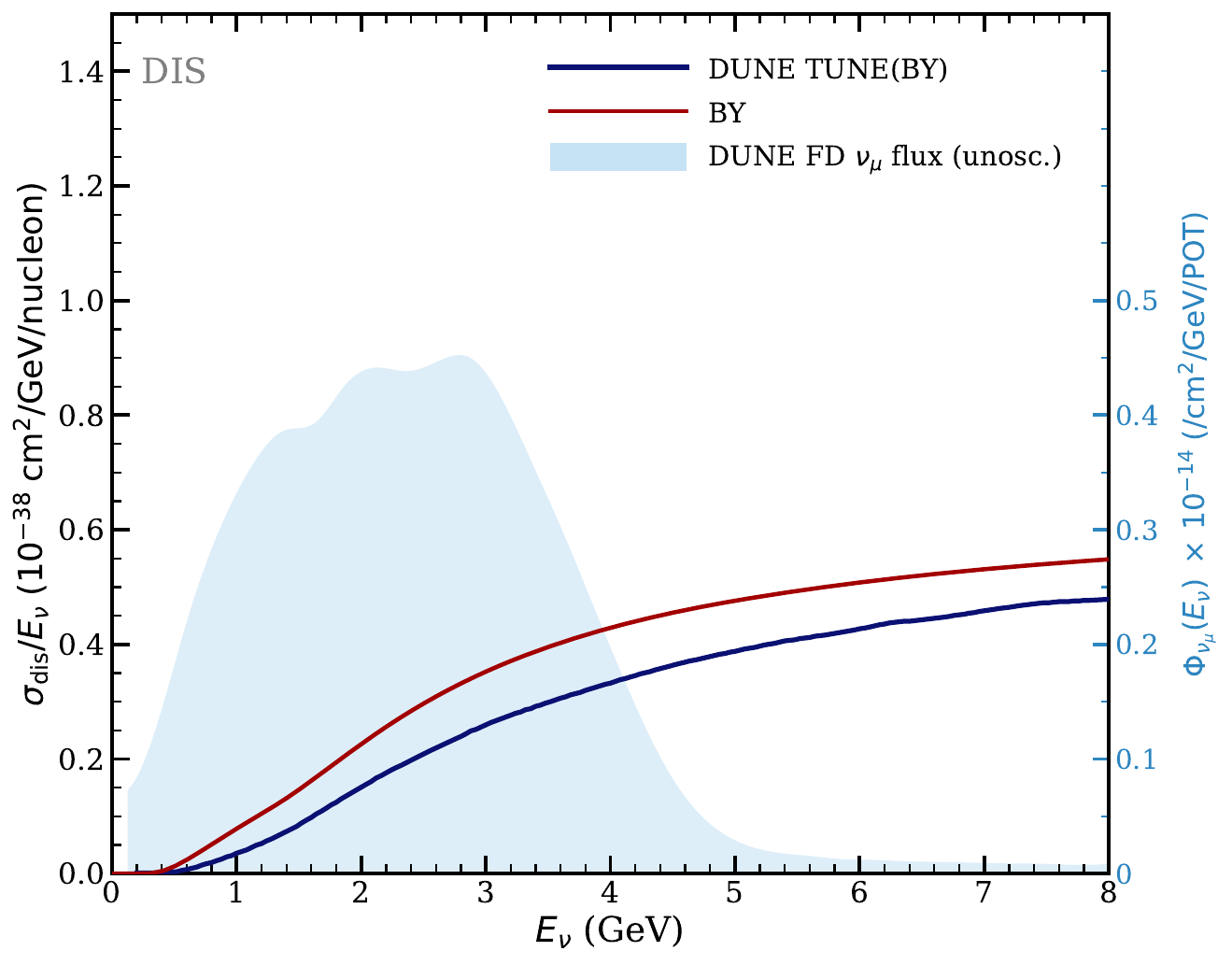}
    \end{minipage}
    \caption{$\nu_{\mu}$--Nucleus($Ar_{40}$) interaction cross section as a function of neutrino energy for total (top left), QEL (top right), RES (bottom left), and DIS (bottom right) channels according to various model combinations in comparison with the DUNE TUNE. The shaded band shows the unoscillated DUNE FD flux (right axis).}
    \label{fig:xsec_model_comp}
\end{figure*}

\section{Results and Discussion}
\label{res}

\subsection{Cross-section comparison}

First, let us discuss the strengths of the different cross-section models that we use in our analysis. In Fig.~\ref{fig:xsec_model_comp}, we illustrate how the different combinations of models affect the $\nu_\mu$ interaction cross-section as a function of neutrino energy. The top-left panel corresponds to the total cross-section, the top-right panel corresponds to the QEL cross-section, the bottom-left panel corresponds to the RES cross-section, and the bottom-right panel corresponds to the DIS cross-section. In each panel, different curves represent different cross-section models. The shape of the DUNE $\nu_\mu$ flux, which demonstrates the relevant energy region, is shown by the shaded blue region in all the panels. 

From the figure, we understand that for the QEL region, the DUNE tune, i.e., the Valencia model, has the strongest strength, followed by HF-CRPA and LS. For RES, the BS model is the best, followed by RS and the DUNE tune. Finally, for the DIS, the BY formalism provides better strength than the DUNE tune. Note that the DUNE tune also uses the RS model for RES and the BY model for DIS. However, this is not the same as our RS and  DIS models due to the different versions of GENIE used by us and the DUNE collaboration, and some specific tuning used in DUNE simulations. The combined effect is visible in the top left panel. In this panel, we see that HF-CRPA+BS has the best strength, followed by HF-CRPA+RS, LS+BS, LS+RS, and finally, the DUNE tune provides the weakest strength in the energy region where the cross-section spectra peaks. However, it is interesting to note that HF-CRPA+RS and LS+BS have similar strengths around 2.5 GeV, where the DUNE flux peaks. From the above discussion, we understand that in the DUNE energy region, the cross-section is mainly governed by the RES, where the cross-section of the DUNE tune is the weakest, and the cross-section for the BS model is best. Therefore, while estimating the DUNE sensitivity, we expect the best sensitivity for the HF-CRPA+BS, while the worst sensitivity for the DUNE tune. We will study this in the next sections.

\subsection{Event spectra comparison}

\begin{figure*}[t]
    \centering

    \begin{minipage}[t]{0.49\textwidth}
        \centering
        \includegraphics[width=\textwidth]{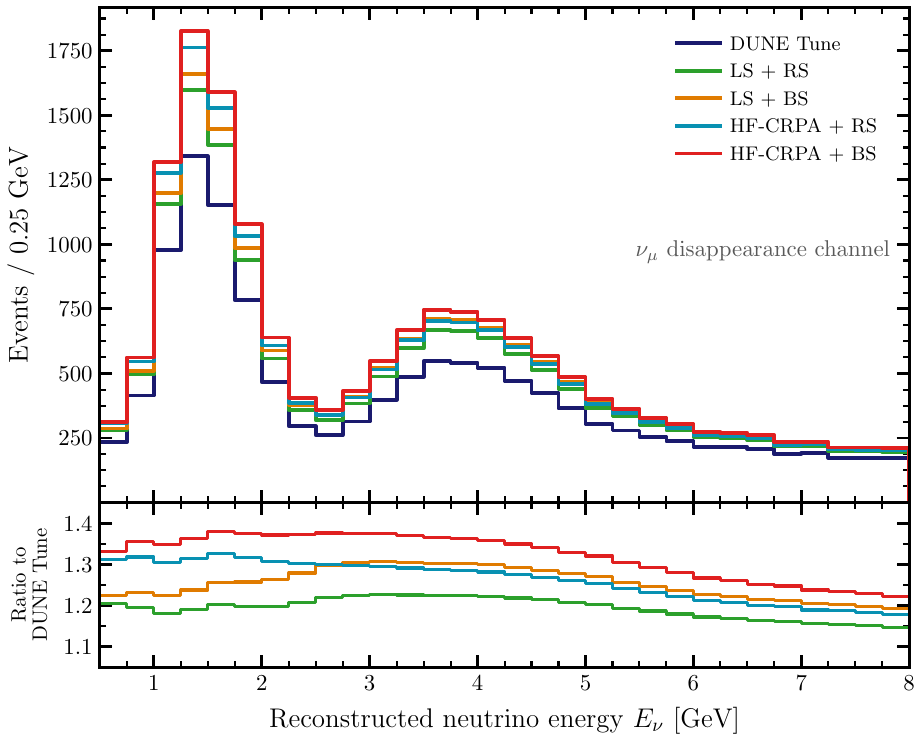}
    \end{minipage}
    \hfill
    \begin{minipage}[t]{0.49\textwidth}
        \centering
        \includegraphics[width=\textwidth]{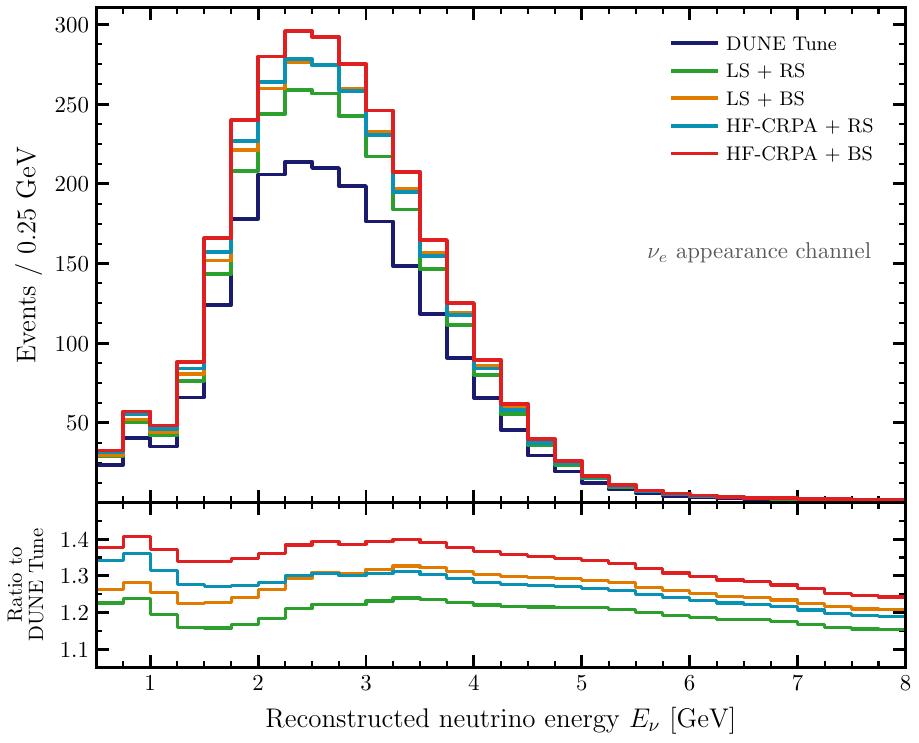}
    \end{minipage}

    \vspace{0.3cm}

    \begin{minipage}[t]{0.49\textwidth}
        \centering
        \includegraphics[width=\textwidth]{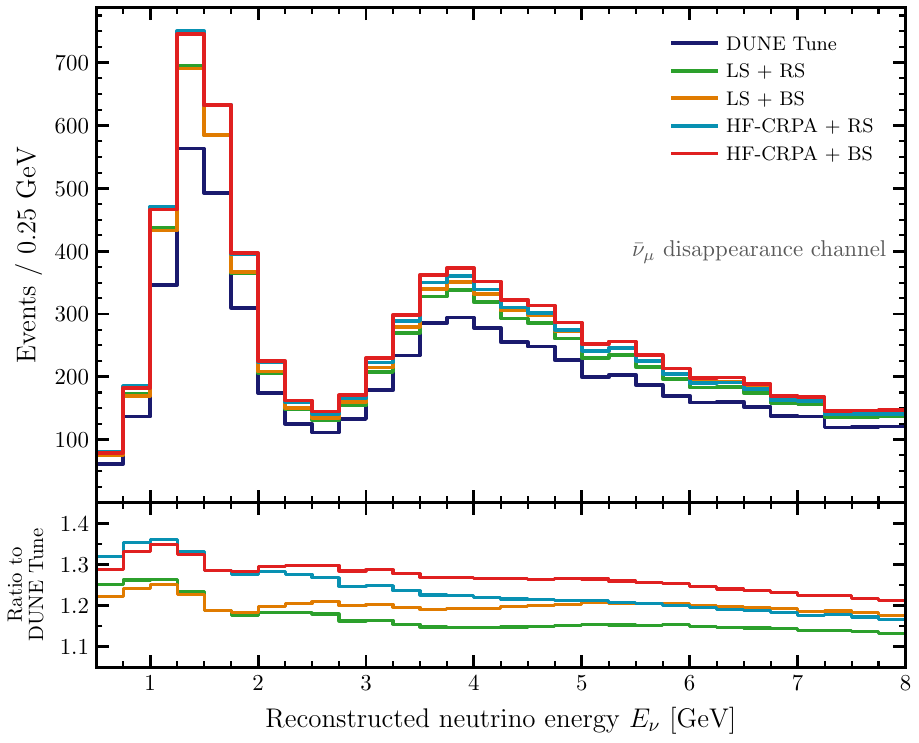}
    \end{minipage}
    \hfill
    \begin{minipage}[t]{0.49\textwidth}
        \centering
        \includegraphics[width=\textwidth]{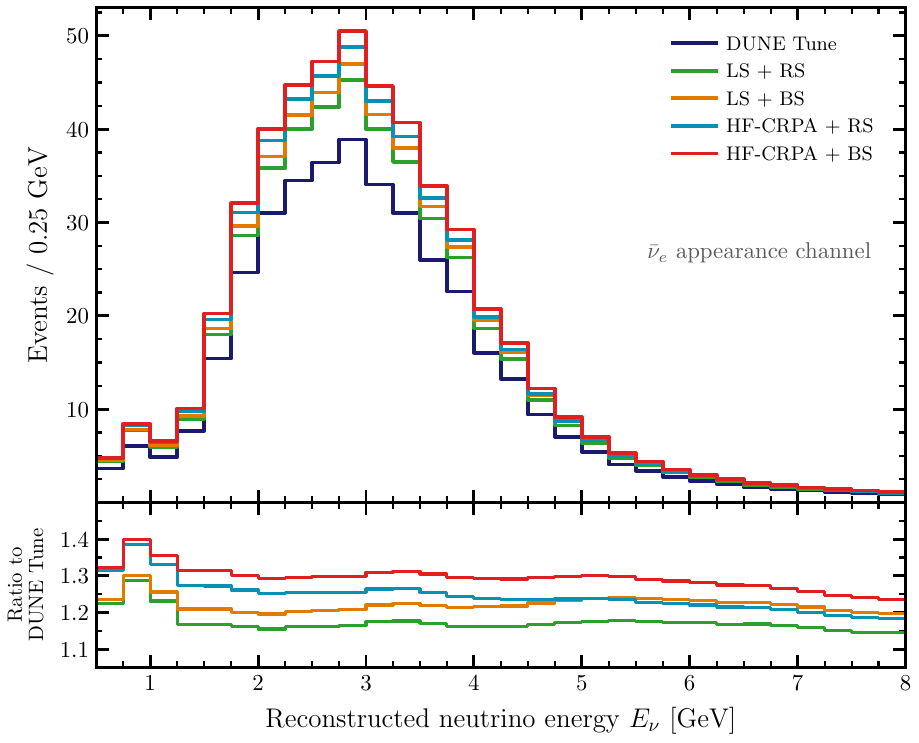}
    \end{minipage}
\caption{Reconstructed far detector event spectra for the four oscillation channels considered in this work: $\nu_\mu$ disappearance (top left), $\nu_e$ appearance (top right), $\bar{\nu}_\mu$ disappearance (bottom left), and $\bar{\nu}_e$ appearance (bottom right). The predictions obtained with the DUNE tune, LS+RS, LS+BS, HF-CRPA+RS, and HF-CRPA+BS interaction model combinations are compared. The lower panels in each subplot show the ratio of the predicted event rates relative to the DUNE tune.}
    \label{fig:event_spectra}
\end{figure*}

\begin{table*}[t]
\centering
\caption{Integrated signal yields in the reconstructed energy range of $0.5$--$8.0$~GeV for the DUNE Tune and the HF-CRPA+BS configuration. The last column gives the increase in the signal event yield relative to the DUNE tune.}
\label{tab:event_summary}
\begin{tabular}{llccccc}
\hline\hline
Beam Mode & Channel & Model & Signal & Increase (\%) \\
\hline

Neutrino (FHC) &
$\nu_e$ appearance &
DUNE Tune &
2037.29 &
-- \\

&
&
HF-CRPA+BS &
2798.94 &
37.4 \\

\hline

Neutrino (FHC) &
$\nu_\mu$ disappearance &
DUNE Tune &
12602.22 &
-- \\

&
&
HF-CRPA+BS &
16923.94 &
34.3 \\

\hline

Antineutrino (RHC) &
$\bar{\nu}_e$ appearance &
DUNE Tune &
389.83 &
-- \\

&
&
HF-CRPA+BS &
507.48 &
30.2 \\

\hline

Antineutrino (RHC) &
$\bar{\nu}_\mu$ disappearance &
DUNE Tune &
6319.67 &
-- \\

&
&
HF-CRPA+BS &
8062.15 &
27.6 \\
\hline\hline
\end{tabular}
\end{table*}

\begin{figure*}
    \centering
    \begin{minipage}[t]{0.48\textwidth}
        \centering
        \includegraphics[width=\textwidth]{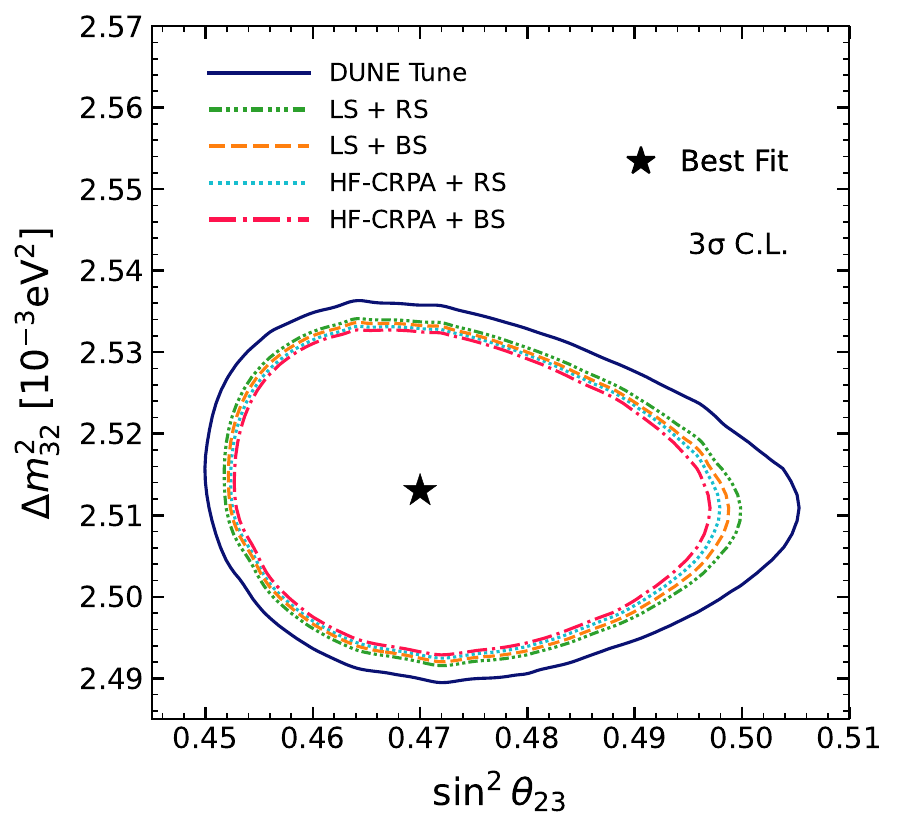}
    \end{minipage}
    \hfill
    \begin{minipage}[t]{0.48\textwidth}
        \centering
        \includegraphics[width=\textwidth]{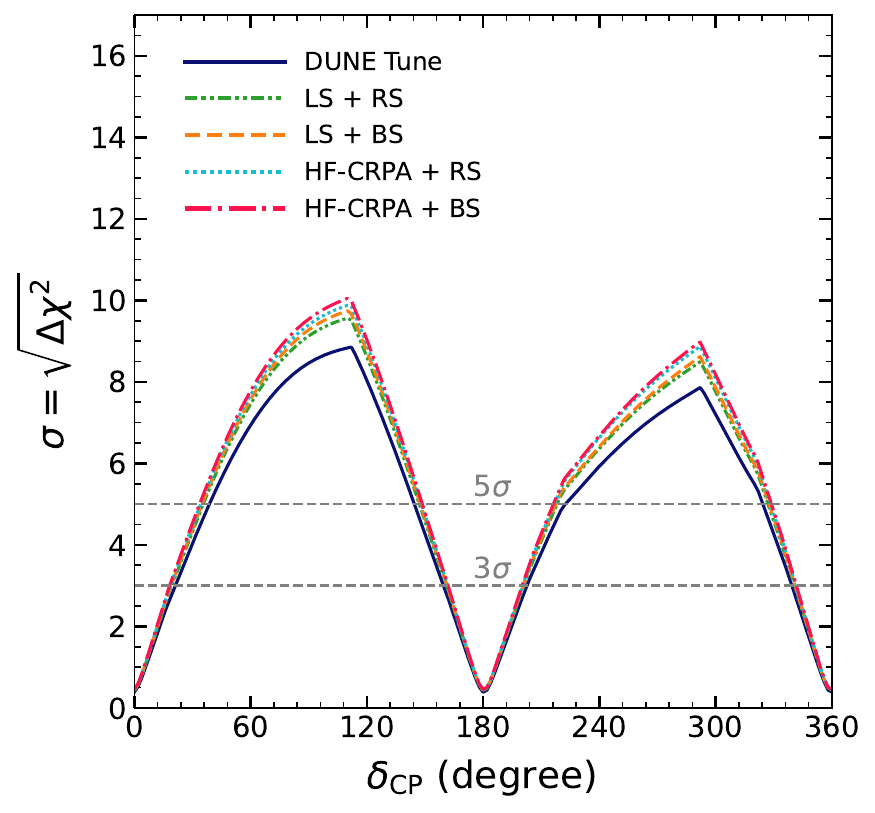}
    \end{minipage}
    \\[0.5em]
    \begin{minipage}[t]{0.48\textwidth}
        \centering
        \includegraphics[width=\textwidth]{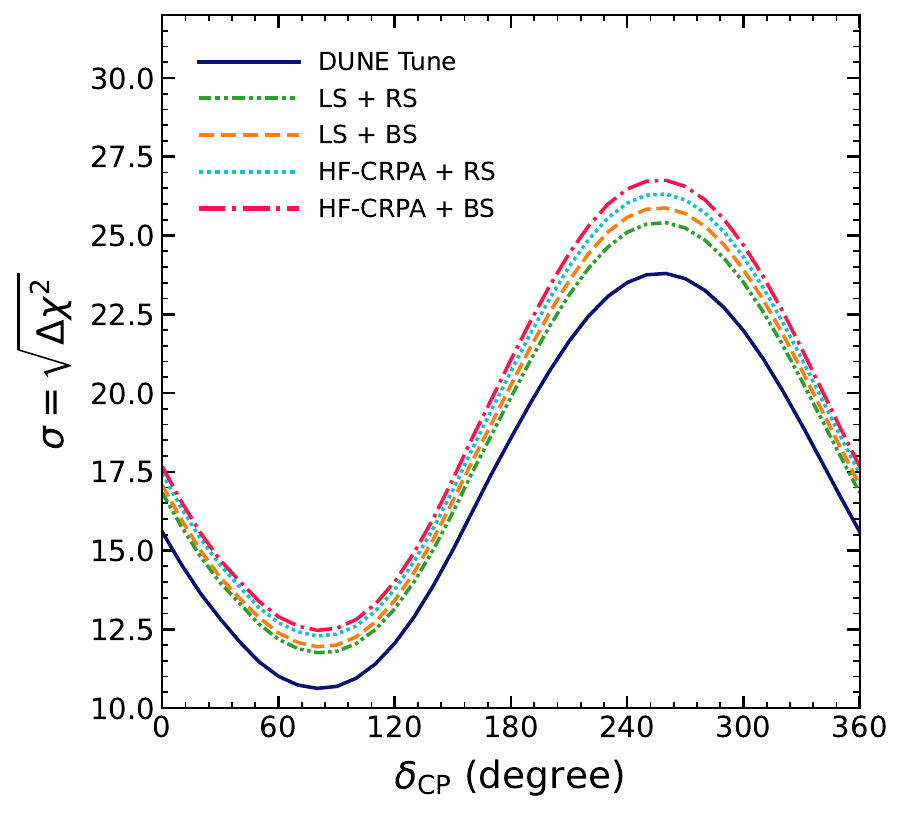}
    \end{minipage}
    \hfill
    \begin{minipage}[t]{0.48\textwidth}
        \centering
        \includegraphics[width=\textwidth]{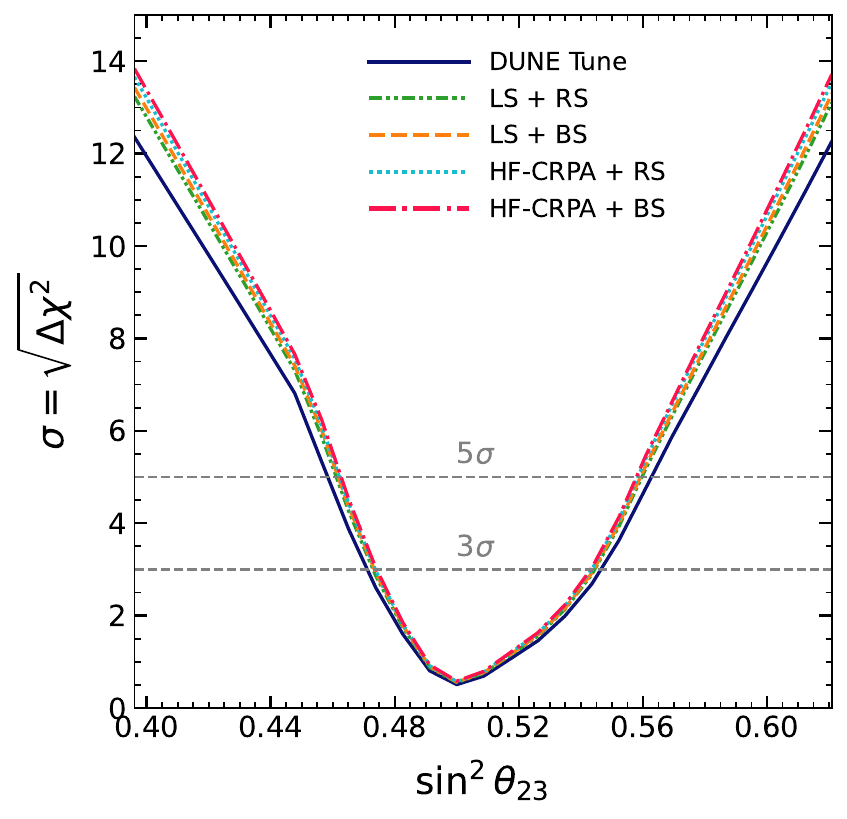}
    \end{minipage}
    \caption{Comparison of DUNE sensitivities to neutrino oscillation parameters obtained using different cross-section model combinations. The upper-left panel shows the $3\sigma$ confidence region in the $(\sin^2\theta_{23},\,\Delta m^2_{32})$ plane, including the best-fit point. The upper-right panel presents the sensitivity to leptonic CP violation as a function of $\delta_{\rm CP}$. The lower-left panel shows the mass ordering sensitivity, while the lower-right panel displays the $\theta_{23}$ octant sensitivity.}
    \label{fig:side_by_side}
\end{figure*}

\begin{table*}
\centering
\caption{Comparison of the oscillation sensitivities obtained with the DUNE tune and the HF-CRPA+BS configuration at the parameter values. The improvement is calculated relative to the DUNE tune.}
\label{tab:improvement_summary}
\begin{tabular}{lcccc}
\hline\hline
Observable & Parameter value & $\Delta\chi^2$ (DUNE Tune) & $\Delta\chi^2$ (HF-CRPA+BS) & Improvement \\
\hline
CP violation          & $\delta_{\rm CP}\approx110^\circ$ & 78.01 & 100.96 & $29.4\%$ \\
Mass ordering         & $\delta_{\rm CP}\approx270^\circ$ & 558.3 &  705.16 & $26.3\%$ \\
Octant sensitivity    & $\sin^2\theta_{23}=0.62$          & 150.52 &  188.2 & $25.0\%$ \\
Octant sensitivity    & $\sin^2\theta_{23}=0.396$         & 152.75 &  191.53 & $25.4\%$ \\
\hline\hline
\end{tabular}
\end{table*}

\begin{table*}
\centering
\caption{Comparison of the projected $3\sigma$ allowed intervals for the atmospheric oscillation parameters obtained with the DUNE tune and the HF-CRPA+BS configuration. The last column gives the reduction in the interval width relative to the DUNE tune.}
\label{tab:precision_summary}
\begin{tabular}{lccccc}
\hline\hline
Parameter &
DUNE Tune ($3\sigma$ interval) &
Width &
HF-CRPA+BS ($3\sigma$ interval) &
Width &
Reduction (\%) \\
\hline

$\sin^2\theta_{23}$ &
$[\,0.44995,\;0.50532\,]$ &
0.05537 &
$[\,0.45268,\;0.49702\,]$ &
0.04434 &
19.9 \\

$\Delta m^2_{32}\;(10^{-3}\,\mathrm{eV}^2)$ &
$[\,2.48947,\;2.53633\,]$ &
0.04686 &
$[\,2.49288,\;2.53274\,]$ &
0.03986 &
14.9 \\

\hline\hline
\end{tabular}
\end{table*}

Having established the improved strength in the cross-section of the HF-CRPA+BS configuration, the immediate next step is to check the prediction of the event spectra in the far detector (FD). Fig~\ref{fig:event_spectra} compares the reconstructed FD event spectra for all the configurations for $\nu_{\mu}$ disappearance (left column) and $\nu_e$ appearance channels (right column) for both neutrino mode (top row) and antineutrino mode (bottom row), and compared with the DUNE tune. In the lower panels of each subplot, we show the ratio of the predicted event rates relative to the DUNE tune. 

As discussed in the previous section, generally, HF-CRPA+BS combination exhibits the largest enhancement over a wide range of reconstructed neutrino energy, followed by HF-CRPA+RS, LS+BS, LS+RS, and the DUNE tune. By having a closer look around energy 2.5 GeV, where the DUNE flux peaks, HF-CRPA+RS and LS+BS have similar predictions for event rates in the neutrino mode for both appearance and disappearance channels. This is the direct consequence of the fact that at this particular energy, HF-CRPA+RS and LS+BS have similar strengths in the total cross-section for the neutrinos. However, we notice that in the antineutrino mode, HF-CRPA+RS exhibits more events than LS+BS at 2.5 GeV. Since the oscillation sensitivity is directly related to the predicted event spectra, the larger samples obtained with a particular configuration naturally lead to an increase in the $\chi^2$ values for CP violation discovery, mass ordering determination, and other parameter measurements. We will study this in the next section.

The total integrated events corresponding to HF-CRPA+BS and DUNE configuration are summarized in the table~\ref{tab:event_summary}. From the table, we see that the HF-CRPA+BS model gives around 27\% - 37\% increase in the signal events as compared to the DUNE TDR tune. 

\begin{figure*}
    \centering
    \begin{minipage}[t]{0.48\textwidth}
        \centering
        \includegraphics[width=\textwidth]{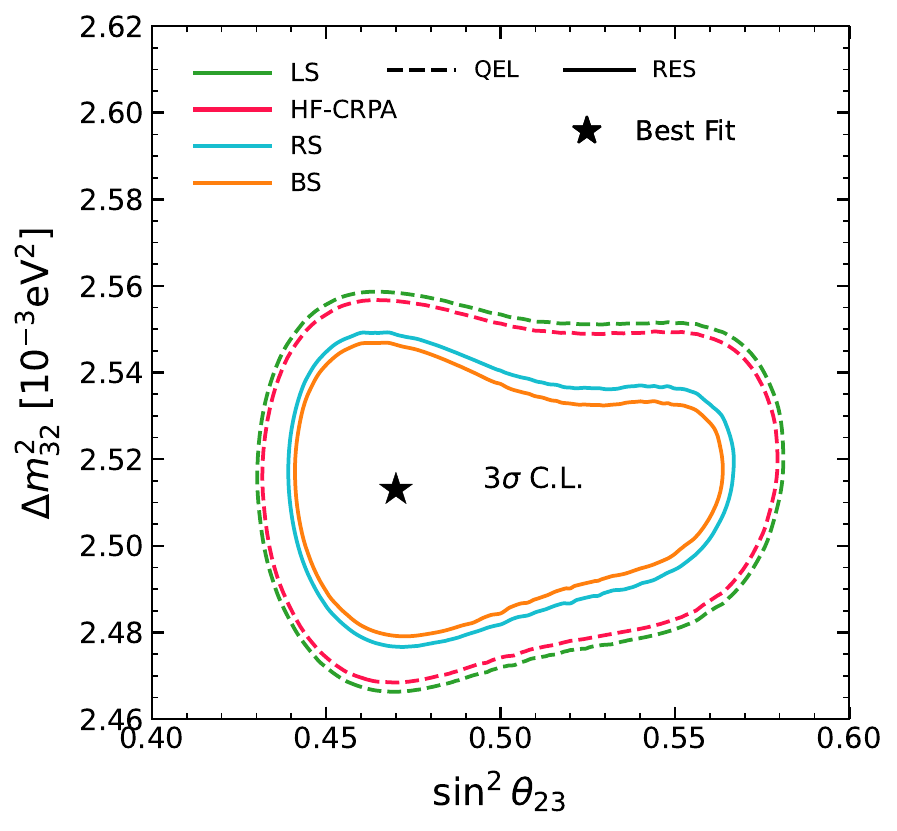}
    \end{minipage}
    \hfill
    \begin{minipage}[t]{0.48\textwidth}
        \centering
        \includegraphics[width=\textwidth]{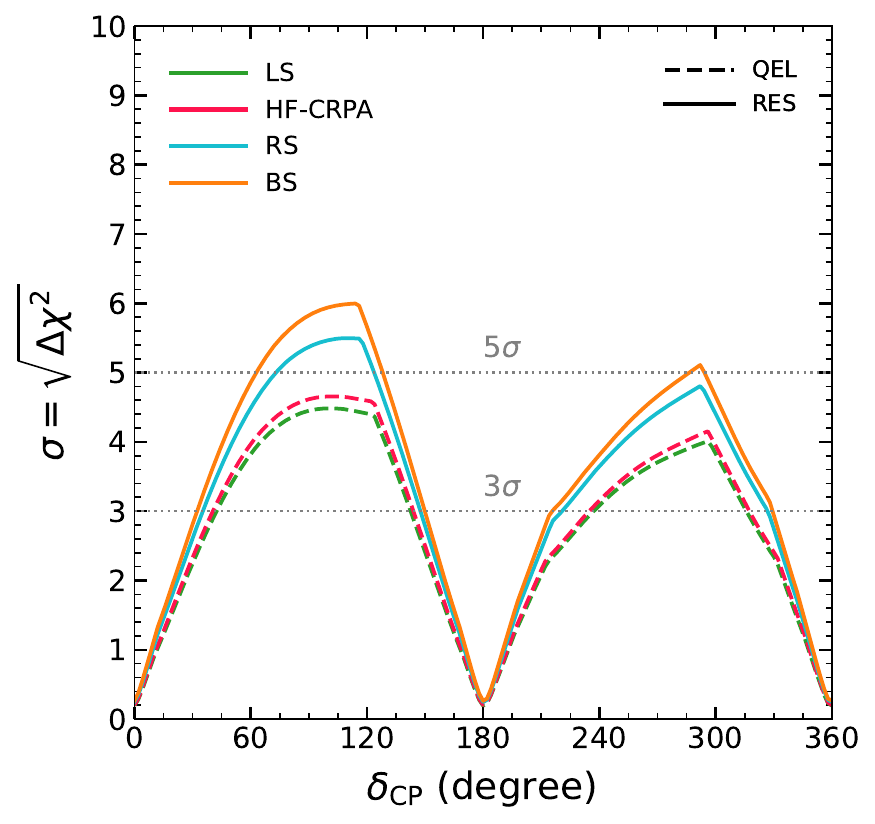}
    \end{minipage}
    \\[0.5em]
    \begin{minipage}[t]{0.48\textwidth}
        \centering
        \includegraphics[width=\textwidth]{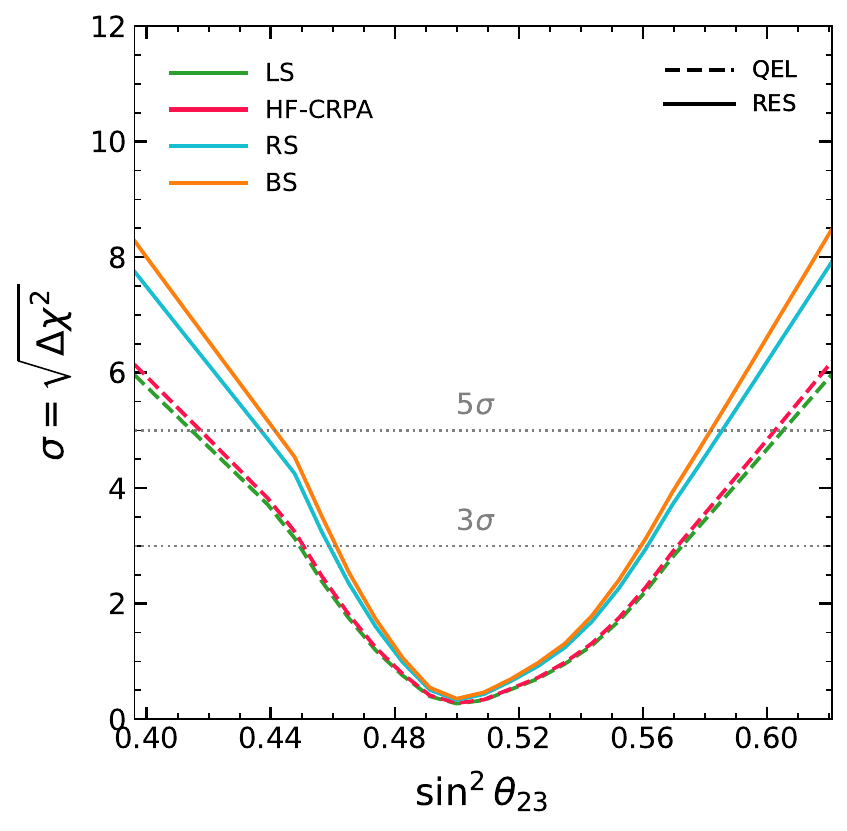}
    \end{minipage}
    \hfill
    \begin{minipage}[t]{0.48\textwidth}
        \centering
        \includegraphics[width=\textwidth]{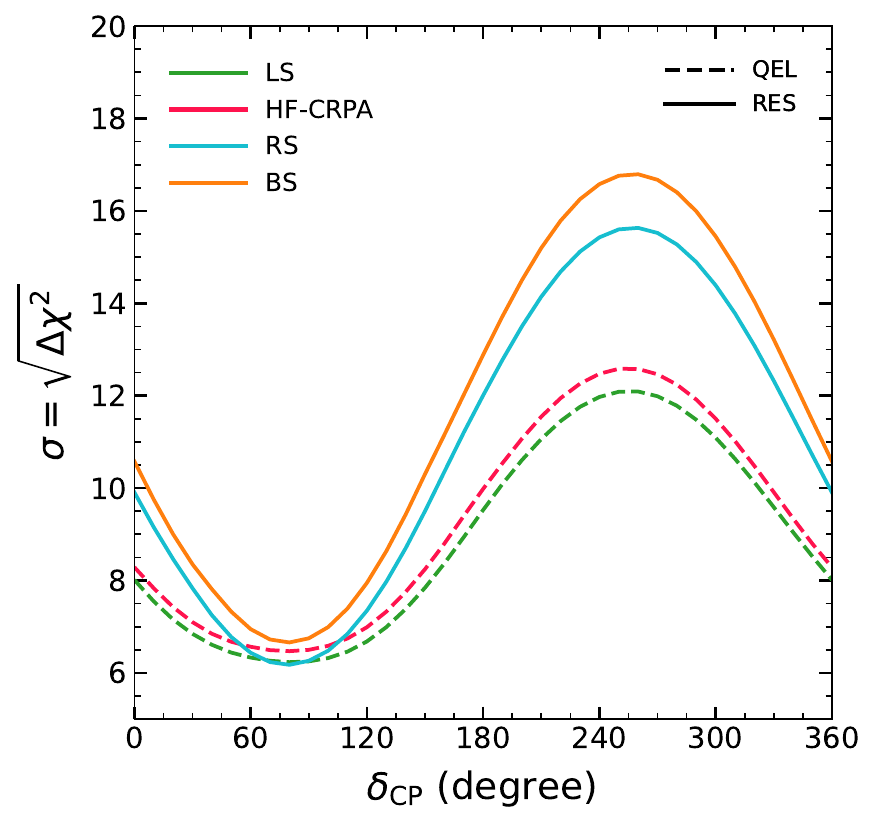}
    \end{minipage}
    \caption{Sensitivities to neutrino oscillation parameters obtained considering only the Quasi-Elastic (QEL) interaction region, comparing the LS and HF-CRPA models, and only the resonance region comparing RS and BS models.}
    \label{fig:side_by_side_combined}
\end{figure*}

\subsection{Sensitivity comparison}

In Fig.~\ref{fig:side_by_side}, we demonstrate the sensitivity of the DUNE experiment with respect to the different cross-section models, considering the total cross-section. The top left panel shows precision of the atmospheric parameters in the $\sin^2\theta_{23} - \Delta m^2_{32}$ plane, the top right shows the CP violation sensitivity as a function of $\delta_{\rm CP}$, the bottom left panel shows the mass ordering sensitivity as a function of $\delta_{\rm CP}$ and the bottom right panel shows the octant sensitivity of $\theta_{23}$ as a function of $\sin^2\theta_{23}$. In each panel different curve corresponds to different cross-section model that we used in our analysis. As predicted in the Figs. \ref{fig:xsec_model_comp} and \ref{fig:event_spectra}, in all the panels, the sensitivities follow exactly the strength of the cross-section. The best sensitivity comes from HF-CRPA+BS, and the worst sensitivity comes from the DUNE tune. In table~\ref{tab:improvement_summary} and~\ref{tab:precision_summary}, we list the change in the sensitivity for the best and the worst cross-section models.

The 3$\sigma$ allowed regions in the $\sin^2\theta_{23} - \Delta m_{32}^2$ plane exhibit a visible contraction when using the HF-CRPA+BS configuration instead of the baseline setup. In particular, the confidence contour is 20\% more precise for $\sin^2\theta_{23}$ and 15\% more precise for $\Delta m^2_{32}$ when we use the HF-CRPA+BS model instead of the baseline DUNE tune. For $\delta_{\rm CP} = 212^\circ$, the CP violation sensitivity increases from $\Delta\chi^2 = 16.161$ for the baseline DUNE configuration to $\Delta\chi^2 = 20.403$ when employing the HF-CRPA + BS framework, corresponding to an enhancement of approximately $26.25\%$. If we consider the maximum CP violating value ($\approx110^\circ$), the corresponding number is $\approx29.5\%$. The sensitivity to the neutrino mass ordering depends strongly on $\delta_{\rm CP}$. This sensitivity is maximized around $\delta_{\rm CP} \simeq 270^\circ$, where CP violating contributions enhance the matter-driven separation between the normal and inverted orderings. Adopting the HF-CRPA+BS configuration increases the peak hierarchy sensitivity by roughly $26.3\%$. The octant sensitivity exhibits the characteristic V-shaped dependence on $\sin^2\theta_{23}$. Near maximal mixing ($\sin^2\theta_{23}=0.5$), the distinction between the lower and upper octants is weakest, causing the sensitivity to approach zero. As the true value departs from maximal mixing, the contrast between the two octant hypotheses becomes increasingly significant, resulting in a monotonic rise in $\Delta\chi^2$. The enhancement becomes more pronounced as the true value moves away from maximal mixing, where the intrinsic octant separation increases, i.e., around $\approx25\%$ at the edges of the x-axis. Near $\sin^2\theta_{23}=0.5$, the sensitivity remains small for all model combinations, as expected.

The above results show that nuclear modeling in the QE–RES transition region can modify DUNE’s oscillation sensitivities by roughly 15–30\%. The enhancements observed in CP violation, the mass ordering, octant, and precision of the atmospheric parameters are all of similar size, indicating that the influence of nuclear modeling is not restricted to any single parameter but instead alters the total oscillation information content of the experiment. In particular, the consistently enhanced performance achieved with the HF–CRPA+BS setup underscores the need for precise modeling of QE and RES interactions in the few-GeV energy range, where the DUNE flux peaks and matter effects are most pronounced.

Next to demonstrate the effect of the individual component of the cross-sections on the estimation of the sensitivity, in Fig.~\ref{fig:side_by_side_combined}, we plotted the same as Fig.~\ref{fig:side_by_side} but only for QEL and RES. The dashed curves are for QE, and the solid curves are for RES. In this figure, we have not shown the sensitivity for the DUNE tune, as we don't have access to the individual component of the cross-section other than the $\nu_\mu$ corresponding to the DUNE TDR. As seen in Fig.~\ref{fig:xsec_model_comp}, we obtain the best sensitivity for BS for the RES and HF-CRPA for QE in all four panels.  

\section{Summary and Conclusions}
\label{conclusion}

In this study, we investigated the impact of cross-section models in determining the physics sensitivities in the standard three-flavor oscillation scenario in DUNE. In particular, we study how different combinations of quasi-elastic and resonance cross-section models affect the extraction of neutrino oscillation parameters at DUNE. All other interaction channels were fixed to match the configuration of the DUNE Technical Design Report, and only the QEL and RES descriptions were altered. This strategy allowed us to isolate the role of nuclear modeling in shaping the oscillation sensitivities. In the QEL sector, we examined the Llewellyn–Smith, and HF–CRPA approaches, while in the resonance region, we employed the Rein–Sehgal and Berger–Sehgal formulations.

Our analysis shows that the selection of cross-section models leads to noticeable shifts in the event spectra and inferred oscillation observables. We have shown that in the DUNE energy region, the cross-section is dominated by RES. Though the DUNE tune for the QEL (Valencia model) has the highest strength, the total cross-section of the DUNE tune is the weakest, as it considers the RS model, and the best strength comes from the HF-CRPA+BS model. A comparative analysis among the models compared in our analysis showed that after the HF-CRPA+BS model, the strengths are in the decreasing order in HF-CRPA+RS, LS+BS, and LS+RS in the energy region where the cross-section spectra peaks. However, it is interesting to note that HF-CRPA+RS and LS+BS have similar strengths around 2.5 GeV, where the DUNE flux peaks. Similar features have also been observed in the event spectra. HF–CRPA+BS combination exhibits the largest enhancement over a wide range of reconstructed neutrino energy, followed by HF-CRPA+RS, LS+BS, LS+RS, and
the DUNE tune. While estimating the sensitivity, we have shown that the allowed contours in the $\sin^{2}\theta_{23} - \Delta m^{2}_{32}$ parameter space contract for certain model choices, signaling improved joint precision on these parameters. Quantitatively, when the HF–CRPA model is used for QEL interactions in combination with the Berger–Sehgal description for RES processes, the sensitivities to CP violation and the mass ordering increase by about $25$–$30\%$ compared to the baseline configuration of DUNE tune, and the octant sensitivity similarly improves away from maximal mixing. 

These results demonstrate that nuclear modeling in the few-GeV quasi-elastic–resonance (QEL–RES) regime can exert a substantial impact on the oscillation sensitivity of DUNE. The associated variations are consistent across multiple key observables, underscoring that accurate and systematically validated cross-section descriptions are essential for achieving precision measurements at forthcoming long-baseline neutrino experiments. Consequently, sustained advancement of nuclear interaction models, along with their rigorous and internally consistent implementation within oscillation analysis frameworks, will be required to fully exploit the physics reach of DUNE. Finally, we would like to say that large differences in the sensitivity due to different cross-section models, i.e., the lack of knowledge of the actual theory, lead to large systematic errors when fitting the actual data, and therefore our study indirectly provides an estimation of current uncertainties on the systematic errors due to the cross-section.

We further reiterate that this work represents the views of the authors and not the DUNE Collaboration.

\acknowledgments

The authors acknowledge the support of the National Institute of Science Education and Research Bhubaneswar  (NISER), and Homi Bhabha National Institute (HBNI), Mumbai. We thank Dr. Stephen Dolan and R.K. Pradhan for some useful discussions on GENIE and HF-CRPA model. MG would like to thank Leon Halić for discussion regarding DUNE tune. This work has been funded in part by the Ministry of Science and Education of the Republic of Croatia, grant No. PK.1.1.10.0002, Swiss National Science Foundation (SNSF) and Croatian Science Foundation (HRZZ) under grant MAPS IZ11Z0$\_$230193 and European Union under the NextGenerationEU Programme. Views and opinions expressed are, however, those of the author(s) only and do not necessarily reflect those of the European Union. Neither the European Union nor the granting authority can be held responsible for them.


\bibliography{mybib}

\end{document}